\documentclass[traditabstract]{aa}
\usepackage{graphicx}
\usepackage{natbib}
\usepackage{txfonts}
\usepackage{subfigure}

\begin{document}
   \title{Evidence of a SiO collimated outflow from a massive YSO in  IRAS\,17233--3606}
   \author{S. Leurini
          \inst{1}
          \and C. Codella\inst{2}
          \and A. Gusdorf\inst{3}
	  \and L. Zapata\inst{4}
          \and A. G\'omez-Ruiz\inst{2}
          \and L. Testi\inst{5,2}          
	  \and T. Pillai\inst{6}
          }

   \offprints{S. Leurini}

\institute{Max-Planck-Institut f\"ur Radioastronomie, Auf dem H\"ugel 69, 53121 Bonn, Germany\\
\email{sleurini@mpifr.de}
       \and INAF - Osservatorio Astrofisico di Arcetri, Largo E. Fermi 5, 50125 Firenze, Italy
       \and LERMA, UMR 8112 du CNRS, Observatoire de Paris, \'Ecole Normale Sup\'erieure, 24 rue Lhomond, F75231 Paris Cedex 05, France
       \and Centro de Radioastronom\'ia y Astrof\'isica, Universidad Nacional Aut\'onoma de M\'exico, Morelia 58090, M\'exico 
        \and ESO, Karl-Schwarzschild Strasse 2, 85748 Garching-bei-M\"unchen, Germany
        \and Caltech, 1200 E. California Blvd, Pasadena, CA 91125, USA
            }

   \date{\today}

  \abstract
   {Studies of molecular outflows in high-mass young stellar objects reveal important information about 
the formation process of massive stars. We therefore selected the close-by IRAS\,17233--3606 massive star-forming region 
to perform SiO observations with the SMA interferometer in the (5--4) line and with the APEX single-dish telescope in the (5--4) and (8--7) transitions.
 In this paper, we present a study of one of the outflows in the region, OF1, which shows several properties similar to jets driven by low-mass protostars, such as HH211 and HH212.
 It is compact and collimated, and  associated with extremely high velocity CO emission, and SiO emission at high velocities. 
We used a state-of-the-art shock model to  
constrain the pre-shock density and shock velocity of OF1. The model 
also  allowed us to self-consistently estimate the mass of the OF1 outflow. The shock parameters inferred by the SiO modelling are comparable with those found for low-mass protostars, only 
with higher pre-shock density values, yielding an outflow mass in agreement with those obtained for molecular outflows driven by early B-type young stellar objects. Our study shows that it is possible to model the SiO emission in high-mass star-forming regions in the same way as for shocks from low-mass young stellar objects.}

   \maketitle
%

\section{Introduction}

Two main theoretical scenarios, based on accretion, are usually proposed
to explain the formation of O-B type stars:
1) the core accretion model \citep{2002Natur.416...59M,2003ApJ...585..850M}, where massive stars form from 
massive cores, and  2) the competitive
accretion model \citep{2007prpl.conf..149B}, where a molecular
cloud fragments into low-mass cores, which form
stars that compete to accrete mass from a common gas reservoir.
Both models predict the existence of accretion disks
around massive young stellar objects (YSOs), and jets driving molecular outflows.
The core accretion model
is a scaled-up scenario of low-mass star formation: in this case,
collimated jets and accretion disk should have the same properties
as in low-mass YSOs.
The competitive accretion
model suggests that massive stars form in
densely clustered environments and that disks and collimated jets
are perturbed by interaction with stellar companions.
To distinguish between these models, it would be helpful to obtain
observation of disk/jet systems in a statistical sample of massive YSOs
 to assess the fraction 
of sources without such system, 
and to derive their properties to compare with low-mass YSOs.  Unlike 
the still-challenging direct observation of
circumstellar disks, outflows/jets extend on larger
scales and are easier to observe. 

High-angular resolution studies of low-mass star-forming regions show that 
silicon monoxide (SiO) thermal emission
is the best tool for tracing jets:
unlike other tracers such as CO, it suffers
minimal contamination from swept-up cavities, and it allows one to probe
obscured regions close to protostars \citep[e.g.][]{2007A&A...462L..53C}.
The formation of SiO can be attributed to
the sputtering of Si atoms from charged grains in a magnetised C-shock
with velocities $\ge$25 km s$^{-1}$
\citep{1997A&A...321..293S,Gusdorf081}.  Whether SiO emission originates 
from shocked ambient material or traces the primary jet itself is still a matter of debate \citep[e.g., ][and references therein]{2007A&A...468L..29C,2012A&A...548L...2C}.
While SiO is extensively observed in low-mass YSOs,  high-angular resolution studies of SiO thermal emission in high-mass star-forming regions
are still limited to a handful
of objects \citep{1999A&A...345..949C,1999AJ....118..477H,2007ApJ...654..361Q,2007A&A...470..269Z,2009ApJ...698.1422Z}
with a typical resolution of several thousand AU.

The prominent far-IR source
IRAS\,17233$-$3606  is one of the best
laboratories for studing massive star formation
because of its close distance \citep[1~kpc,][]{2011A&A...530A..12L}, high luminosity
($\sim1.7 \times 10^4 L_\odot$), and  relatively simple geometry.
In a previous study in CO with the SMA \citep[][hereafter Paper~I]{2009A&A...507.1443L},
we resolved multiple outflows with high collimation factors, extremely high velocity (EHV) emission
($|\varv-\varv_{\rm {LSR}}|$ up to 200~km~s$^{-1}$), and flow parameters typical of massive YSOs. 
However, the severe contamination from swept-up components and the very
elongated beam ($\sim 5\arcsec\times3\arcsec$) did not allow  us to understand the number of outflows
in the region or to identify the corresponding driving sources and 
to assess 
the occurrence of jets. 

In this paper, we present SMA SiO(5--4)  and complementary APEX SiO(5--4) and 
(8--7) observations  
aimed at separating the outflow multiplicity in the cluster. 
We  focus on the OF1 flow, which 
is associated with compact jet-like H$_2$ emission at 2.12$\mu$m and extremely high velocity CO, and is
possibly driven by a zero-age main-sequence (ZAMS) B1 star \citep{2008AJ....136.1455Z}.

\section{Observations and data reduction}

\begin{table}
\caption{Summary of the observations.}\label{table_obs}      
\centering
\begin{tabular}{lc}
\hline \hline
\multicolumn{2}{c}{SMA}\\

Centre of the map&$\alpha_{\rm{J2000}}~17^{\rm h}26^{\rm m}42\fs651$\\        
&$\delta_{\rm{J2000}}~-36\degr09\arcmin38\farcs00$\\
Beam size& $3\farcs2 \times 2\farcs5$, P.A. 16$\degr$\\
Central frequency&217104.980~MHz\\
Velocity resolution& 2 km~s$^{-1}$\\
Bandwidth& 2~GHz\\
\hline
\multicolumn{2}{c}{APEX}\\

Centre of the map&$\alpha_{\rm{J2000}}~17^{\rm h}26^{\rm m}42\fs930$\\        
&$\delta_{\rm{J2000}}~-36\degr09\arcmin20\farcs00$\\
Beam size&30\farcs5\\
Central frequency$^{\mathrm{a}}$&217104.9800~MHz, 347330.5786~MHz\\
Velocity resolution$^{\mathrm{a}}$& 0.3~km~s$^{-1}$, 0.2~km~s$^{-1}$\\
Bandwidth& 1 GHz\\
\hline
\end{tabular}
\begin{list}{}{}
\item[$^{\mathrm{a}}$] for SiO(5--4) and (8--7), respectively
\end{list}
\end{table}

\subsection{APEX}\label{obs_a}
We mapped a region of $70\arcsec \times 70\arcsec$ in the SiO(5--4) line at 217.105~GHz 
and of $36\arcsec \times 36\arcsec$ in SiO(8--7) at  347.331~GHz in raster mode 
with the APEX-1 and -2 receivers, respectively, on the  
APEX\footnote{APEX is a collaboration between
the Max-Planck-Institut f\"ur Radioastronomie, the European Southern
Observatory, and the Onsala Space Observatory.} telescope in May 2010. 
The spacing between pointing centres was chosen to be 14\arcsec at 217~GHz and 9\arcsec at 347~GHz.
The centre of the maps was taken as 
$\alpha_{[\rm{J}2000]}$=$17^{\rm h}26^{\rm m}42\fs93$, $\beta_{[\rm{J}2000]}$=$-36\degr09\arcmin20\arcsec$.
The APEX facility fast Fourier transform spectrometers \citep{2006A&A...454L..29K} allow a simultaneous observations of 1~GHz bandwidth per receiver, centred on the SiO(5--4) and (8--7) lines. The velocity resolution of the original data is $\sim0.3$~km~s$^{-1}$ at 217~GHz
and $\sim0.2$~km~s$^{-1}$ at 347~GHz.
Data were converted into $T_{MB}$ units
assuming a forward efficiency of 0.95 for both receivers, and a beam efficiency of 0.75 and 0.73 
for the SiO(5--4) and (8--7) line, respectively. Pointing was checked on the 
CO(2--1) and (3--2) lines towards
RAFGL4211, NGC6072 and NGC6302
and was found to have an r.m.s. of 
$(3\farcs3,3\farcs8)$ for the APEX-1 data, and of $(1\farcs4,0\farcs6)$ for 
the APEX-2 data. 
Data cubes were made with the XY$\_$MAP task in the GILDAS software\footnote{http://www.iram.fr/IRAMFR/GILDAS} with 
a common angular resolution of 30\farcs5. The final spectral resolution
is 3.0~km~s$^{-1}$, with an r.m.s. of 0.02~K for SiO(8--7) and 0.1~K for SiO(5--4).

\subsection{SMA}\label{obs_sma}
SiO(5--4) line observations of IRAS\,17233--3606 were
obtained in June 2008 with the Submillimeter Array\footnote{The Submillimeter
Array (SMA) is a joint project between the Smithsonian Astrophysical
Observatory and the Academia Sinica Institute of Astronomy and
Astrophysics, and is funded by the Smithsonian Institution and the
Academia Sinica.} (SMA) in its compact-north configuration with 
baselines ranging in projected length from 6 to 74 k$\lambda$.
The central frequency of the lower
sideband receivers was 217.1049 GHz, while the upper sideband 
central frequency was 227.1049 GHz. The full bandwidth of the SMA correlator was 4 GHz (2 GHz in each
band). The SMA digital correlator was
configured 
to provide a spectral 
resolution of 0.406 MHz ($\sim$ 0.56 km s$^{-1}$). The primary beam at 217~GHz has a FWHM diameter of $\sim54''$.

The region was covered with a mosaic of two fields
($\alpha_{[\rm{J}2000]}=17^{\rm h}26^{\rm m}42\fs64$,
$\delta_{[\rm{J}2000]}$=$-$36$^{\circ}$09$'$18$\farcs$0, 
$\alpha_{[\rm{J}2000]}=17^{\rm h}26^{\rm m}42\fs64$,
$\delta_{[\rm{J}2000]}$=$-$36$^{\circ}$09$'$38$\farcs$0).
Bandpass calibration was made with 3C279. We used Titan for
the flux calibration with an accuracy of
15-20$\%$, based on the SMA monitoring of quasars.
The gain  calibration was done via frequent observations of the quasar
 1626-298.
The zenith opacity, measured with the NRAO tipping
radiometer located at the Caltech Submillimeter Observatory, was
very stable at $\sim0.2$, indicating reasonable weather conditions
during the observations.

The initial flagging and calibration was performed with the IDL superset
MIR\footnote{The MIR cookbook by Charlie Qi can be
found at http://cfa-www.harvard.edu/$\sim$cqi/mircook.html.}. The
imaging and data analysis was conducted in MIRIAD
\citep{1995ASPC...77..433S} and
GILDAS.  Images were produced using natural weighting.
The resulting synthesised beam is $3\farcs2 \times 2\farcs5$
with a P.A. of 16$\degr$. 
The final r.m.s. corresponds to 0.1~Jy/beam  with a spectral resolution of 2~km~s$^{-1}$. The conversion factor between Jy/beam and K is 3.4.

\section{Observational results}
\subsection{SMA data}\label{res_sma}

\begin{table}
\caption{Summary of relevant positions in the IRAS\,17233--3606  cluster.}\label{pos}      
\centering
\begin{tabular}{lrr}
\hline \hline
mm-continuum& $\alpha_{\rm[J2000]}=17^{\rm h}26^{\rm m}42\fs51$&$\delta_{\rm[J2000]}=-36\degr09\arcmin18\farcs32$\\
VLA-2a&$\alpha_{\rm[J2000]}=17^{\rm h}26^{\rm m}42\fs51$&$\delta_{\rm[J2000]}=-36\degr09\arcmin18\farcs04$\\
VLA-2b&$\alpha_{\rm[J2000]}=17^{\rm h}26^{\rm m}42\fs55$&$\delta_{\rm[J2000]}=-36\degr09\arcmin17\farcs67$\\
VLA-2c&$\alpha_{\rm[J2000]}=17^{\rm h}26^{\rm m}42\fs61$&$\delta_{\rm[J2000]}=-36\degr09\arcmin17\farcs35$\\
VLA-2d&$\alpha_{\rm[J2000]}=17^{\rm h}26^{\rm m}42\fs69$&$\delta_{\rm[J2000]}=-36\degr09\arcmin17\farcs05$\\
SiO-APEX&$\alpha_{\rm[J2000]}=17^{\rm h}26^{\rm m}41\fs77$&$\delta_{\rm[J2000]}=-36\degr09\arcmin20\farcs00$\\
H$_2$-OF1&$\alpha_{\rm[J2000]}=17^{\rm h}26^{\rm m}42\fs31$&$\delta_{\rm[J2000]}=-36\degr09\arcmin14\farcs49$\\
SiO-HV$^{\mathrm{a}}$&$\alpha_{\rm[J2000]}=17^{\rm h}26^{\rm m}42\fs29$&$\delta_{\rm[J2000]}=-36\degr09\arcmin13\farcs74$\\
SiO-LV$^{\mathrm{b}}$&$\alpha_{\rm[J2000]}=17^{\rm h}26^{\rm m}42\fs40$&$\delta_{\rm[J2000]}=-36\degr09\arcmin15\farcs62$\\
R1$^{\mathrm{c}}$&$\alpha_{\rm[J2000]}=17^{\rm h}26^{\rm m}42\fs15$&$\delta_{\rm[J2000]}=-36\degr09\arcmin13\farcs28$\\
\hline
\end{tabular}
\begin{list}{}{}
\item[$^{\mathrm{a}}$] peak of the HV red-shifted SMA-SiO emission (Fig.~\ref{zoom})
\item[$^{\mathrm{b}}$] peak of the LV red-shifted SMA-SiO emission (Fig.~\ref{zoom})
\item[$^{\mathrm{c}}$] peak of the EHV red-shifted CO emission (see Paper~I and Fig.~\ref{zoom})
\end{list}
\end{table}

\begin{figure*}
\centering
\includegraphics[angle=-90,width=16cm]{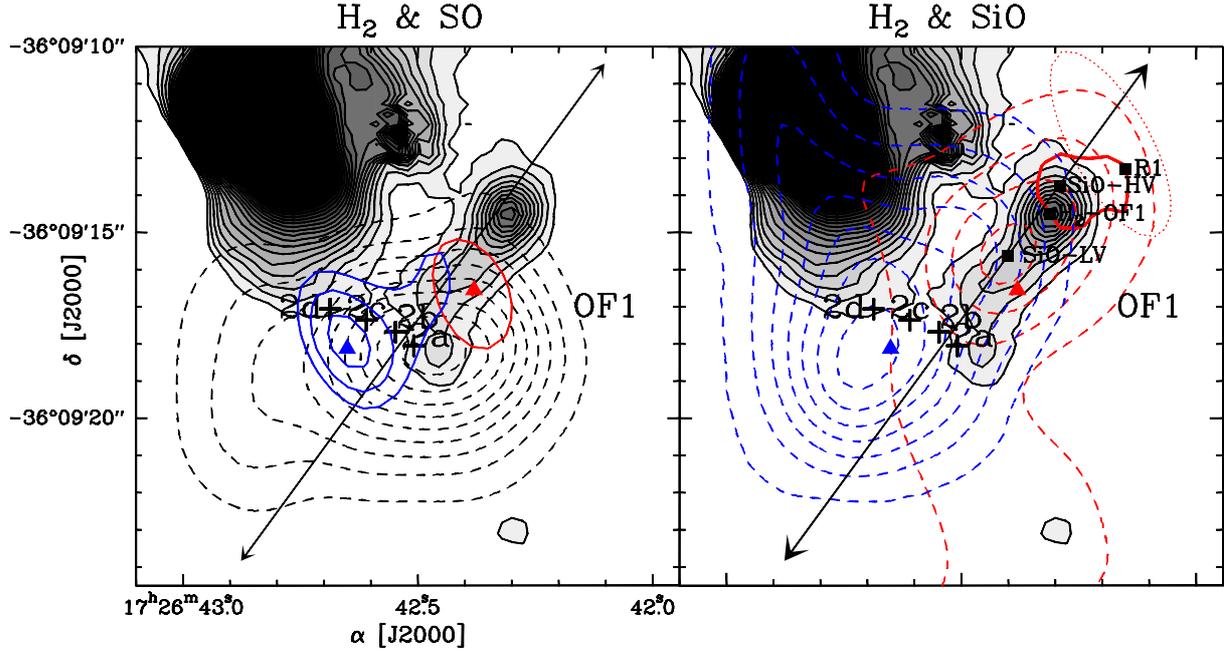}
\caption{{\it{Left:}} Dashed contours (from 10$\sigma$, 0.23~Jy/beam, in step of 10$\sigma$) are the 1.4~mm continuum emission observed with the SMA.   The solid blue and red contours are the SO($5_6-4_5$) blue- and red-shifted emission from Paper~I (blue: from 80 to 104 Jy beam$^{-1}$ in steps of 10 Jy beam$^{-1}$; red: 10 and
12 Jy beam$^{-1}$).
{\it{Right:}} Integrated emission
of the  SiO(5--4) line: the dashed blue and red contours are the LV blue- 
and red-shifted SiO emission
 ($\varv_{\rm{bl}}=[-20,-10]$~km~s$^{-1}$, from 5$\sigma$ (2.5 Jy/beam) in step of 10$\sigma$; 
$\varv_{\rm{rd}}=[+10,+39]$~km~s$^{-1}$, from 5$\sigma$ (4 Jy/beam) in step of 10$\sigma$), 
the thick contour the HV red-shifted emission ($\varv_{\rm{rd}}=[+50,+70]$~km~s$^{-1}$, 4$\sigma$ (2.4 Jy/beam)). 
 The red dotted
contour is the EHV CO(2--1) red-shifted emission (see Fig.~3 of Paper~I).
 The black squares mark the peak positions of the EHV CO(2--1) (R1) red-shifted emission, 
of the SiO(5--4) LV (SiO-LV) and HV red-shifted emission (SiO-HV) , and of the H$_2$ emission in OF1 (H$_2$-OF1).
In both panels, the grey scale and the solid black contours represent  
the H$_2$ emission at 2.12$\mu$m. 
The black crosses mark the positions of the VLA-2a,-2b,-2c and -2d 1.3~cm continuum sources 
\citep{2008AJ....136.1455Z}; 
the arrow marks the OF1 outflow.  The blue and red triangles mark the peaks of the blue- and red-shifted SO($5_6-4_5$) emission.}\label{zoom}
\end{figure*}

The  continuum  emission   of the region at  217~GHz     is  shown  in
Fig.~\ref{zoom}. We performed a 2-D Gaussian fit of the mm continuum emission, and
determined its peak to be at $\alpha_{\rm J2000}=17^{\rm h}26^{\rm m}42\fs506$,
$\delta_{\rm J2000}=-36\degr09'18\farcs319$ with  positional errors on the fit of $\sim0\farcs04$.
The peak flux  is $\sim
2.10\pm0.04$~Jy beam$^{-1}$, the integrated flux  6.69 Jy.  
The results agree very well with the continuum data at 230~GHz presented by 
\citet{2011A&A...530A..12L}.

\begin{figure*}
\centering
\includegraphics[angle=-90,width=12cm]{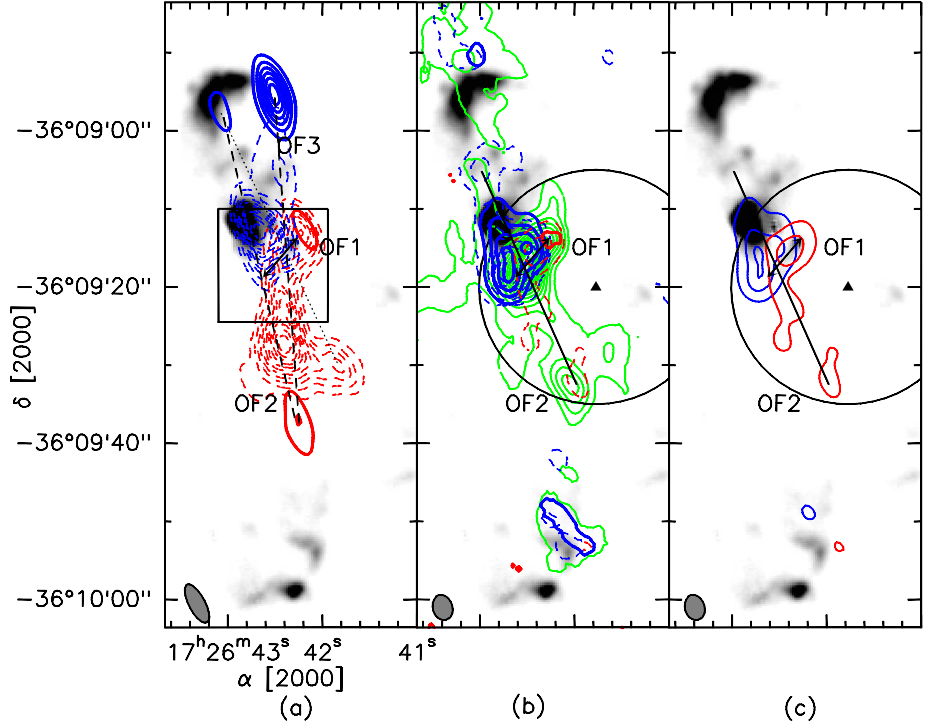}
\caption{{\it Left panel:} Integrated emission of the CO(2--1) blue- and red-shifted wings. The solid
contours show the EHV blue- and red-shifted emission; the dashed contours mark the HV blue- 
and red-shifted emission (see Paper~I, Fig.~3). The dashed and solid lines outline the possible molecular outflows
 (OF1, OF2, OF3); the dotted line marks the alternative direction of the OF2 outflow. 
The central box
is the region mapped in Fig.~\ref{zoom}. {\it Middle panel:} Integrated emission
of the  SiO(5--4) line: the dashed blue and red contours are the LV blue- 
and red-shifted emission
 ($\varv_{\rm{bl}}=[-20,-10]$~km~s$^{-1}$, from 5$\sigma$ (2.5 Jy/beam) in steps of 10$\sigma$; 
$\varv_{\rm{rd}}=[+10,+39]$~km~s$^{-1}$, from 5$\sigma$ (6 Jy/beam) in steps of 10$\sigma$), 
the thick contours the HV blue- and red-shifted emission ($\varv_{\rm{bl}}=[-50,-20]$~km~s$^{-1}$, 
from 5$\sigma$ (4 Jy/beam) in steps of 10$\sigma$; 
$\varv_{\rm{rd}}=[+50,+70]$~km~s$^{-1}$, 4$\sigma$ (2.4 Jy/beam)). 
The green contours are the integrated emission at ambient
velocities ($\varv_{\rm{gr}}=[-10,+10]$~km~s$^{-1}$; from 5$\sigma$ (3 Jy/beam) in steps of 10$\sigma$). 
 {\it Right panel:} Integrated emission
of the  SiO(5--4) line in the velocity ranges used for  modelling  the SiO emission in Sect.~\ref{sio}:
the  blue and red contours are the  blue- 
and red-shifted emission in the ranges
 ($\varv_{\rm{bl}}=[-30,-20]$~km~s$^{-1}$, from 5$\sigma$ (4 Jy/beam) in steps of 10$\sigma$; 
$\varv_{\rm{rd}}=[+10,+39]$~km~s$^{-1}$, from 5$\sigma$ (6.5 Jy/beam) in steps of 10$\sigma$).
In the three panels, the solid line marks the axis of OF2 as identified in SiO.
The arrowed line marks that of OF1. The solid circle represents the APEX 30\farcs5 beam
centred on  $\alpha_{[J2000]}=17^{\rm h}26^{\rm m}41\fs77$, $\delta_{[J2000]}=-36\degr09\arcmin20\arcsec$, the position used in Sect.~\ref{sio} for our modelling of the SiO emission (SiO-APEX, see Table~\ref{pos}) and here labelled by a black triangle.
 In both panels, the grey scale represents  
the H$_2$ emission at 2.12$\mu$m (Paper I).}
\label{overview}
\end{figure*}

Figure~\ref{overview} shows the  emission of the SiO(5--4) line observed with the SMA and integrated over three different velocity ranges: the ambient velocity  and the blue- and red-shifted velocities, which include emission at low (LV) and high velocities (HV).
The systemic velocity of the source is $-3.4$ km~s$^{-1}$ \citep{1996A&AS..115...81B}. 
However large uncertainties in the determination of the ambient velocity were reported in previous SMA observations of complex molecules towards the peak of the millimetre continuum emission
\citep{2011A&A...533A..85L}. Therefore, 
 we conservatively assumed ambient emission (in green in Fig.~\ref{overview}) 
from -10 to 10 km~s$^{-1}$.
Blue- and red-shifted emission is clearly detected towards at least 
two of the outflows identified in the region in Paper I, 
OF1 and OF2.  SiO emission is also 
detected near the H$_2$ emission to the south of the IRAS\,17233--3606  cluster, 
thus supporting the scenario 
that these H$_2$ structures are also associated with the outflows, as there is no H$_2$ emission elsewhere in the map.
Ambient emission is detected all along OF2 and
towards the H$_2$ bow to the north of IRAS\,17233--3606 . 
While in our previous study we suggested two possible axes
for OF2 (Fig.~\ref{overview}, left panel), the SiO(5--4) data solve the 
degeneracy and clearly outline its axis (Fig.~\ref{overview}, right panel).  

\begin{figure}
\centering
 \includegraphics[angle=-90,width=6cm]{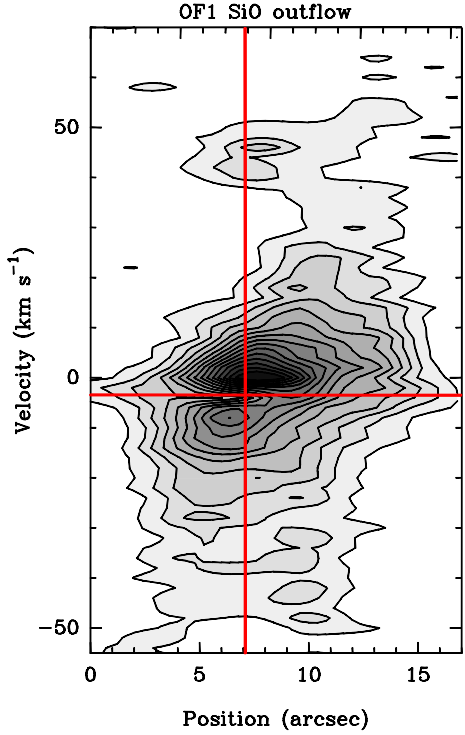}
\caption{Position-velocity diagram of the SiO(5--4) transition (SMA data) along the OF1 axis shown in Fig.~\ref{zoom}. The vertical line
marks the mm continuum peak, 
the horizontal line the ambient velocity (-3.4~km~s$^{-1}$, \citealt{1996A&AS..115...81B}). 
Contours are from $2\sigma$ (0.2~Jy/beam) in steps of $4\sigma$.}\label{pv}
\end{figure}

 The blue- and red-shifted SiO emission shown in Fig.~\ref{overview} concentrates along the axis of OF1 (see Fig.~\ref{zoom}).
The OF1 outflow shows interesting properties that deserve a closer look (Fig.~\ref{zoom}). 
 The position-velocity diagram of the SiO(5-4) line 
along the axis of OF1 is presented in Fig.~\ref{pv}: the SiO emission follows a 
linear velocity-distance relation, typically referred to as the
Hubble-law, where the maximum radial velocity is proportional
to position \citep[e.g., ][]{1996ApJ...459..638L}.
The morphology of OF1 as traced by SiO is very similar to that pictured by H$_2$. 
Its red lobe, observed in H$_2$ and  in EHV  CO emission  (up to $\sim 120$~km~s$^{-1}$), is here  
confirmed by SiO red-shifted emission at LV and HV velocities  
(up to $\sim55$~km~s$^{-1}$ at a $3\sigma$ level, see Fig.~\ref{pv}). 
Interestingly, the HV SiO emission  
peaks between the apex of the H$_2$ emission and the peak of the EHV CO emission. 
The blue lobe is traced by LV and HV CO and SiO (see Fig.~\ref{zoom}). The LV SiO emission is also very similar to that of SO($5_6-4_5$) (see Fig.~\ref{zoom}, left panel), which is the only tracer, together with SiO, that clearly shows a bipolar morphology along OF1. Interestingly, the similarity between the SiO and SO emissions follows the recent result of \citet{2010A&A...522A..91T}, who found that the SiO and SO molecules have very similar line profiles in  collimated molecular outflows driven by Class 0 low-mass sources.
Based on the FWHM of the LV blue- and red-shifted SiO integrated emission, 
we infer an observed area of 
$\sim30''$ squared  and a length of $\sim9\farcs6$, corresponding to a deconvolved area of $22''$ squared and to a deconvolved length of $\sim9\farcs0$. However, the emission is unresolved along the minor axis of OF1.

 It is interesting to compare the SiO and H$_2$ images of the OF1 outflow
with the 
recent findings towards typical low-mass protostars such
as HH211 and HH212 \citep[see e.g., ][]{2006ApJ...636L.141H,2007A&A...462L..53C}
located at $\sim$ 300--500~pc distance.
The SiO emission in HH211 and HH212 is confined to a highly collimated
bipolar jet, well
detected thanks to sub-arcsec spatial resolution of the SMA and Plateau de Bure images.
The SiO jet can be
traced  to within 500--1000 AU of the protostar, in a region that is
heavily obscured
in H$_2$ images. On the other hand, the brightest H$_2$
knots are located in an outer region and lack a well-defined SiO
counterpart, probably
tracing more powerful shock where
SiO does not form/survive or where the SiO excitation is extremely high
and 
consequently the low-J mm-transitions are weak.
In addition, the position-velocity diagrams suggest that the SiO emission increases in 
velocity
with increasing distance from the driving source.
The present OF1 images, although observed with a poorer spatial
resolution, seem to
suggest a similar picture: (i) the position of the SiO red lobe is not
correlated with
the brightest H$_2$ knot, and (ii) the SiO-LV emission peaks closer to
the driving source
than the SiO-HV peak. Only future higher spatial resolution
observations,
possibly with ALMA, will allow a proper characterisation of the region.

 Comparisons between the SMA SiO(5--4) flux and the flux observed with the APEX telescope in the same transition are discussed in the next section.

\subsection{APEX data}\label{res_a}
\begin{figure}
\centering
\subfigure{\includegraphics[angle=-90,width=8cm]{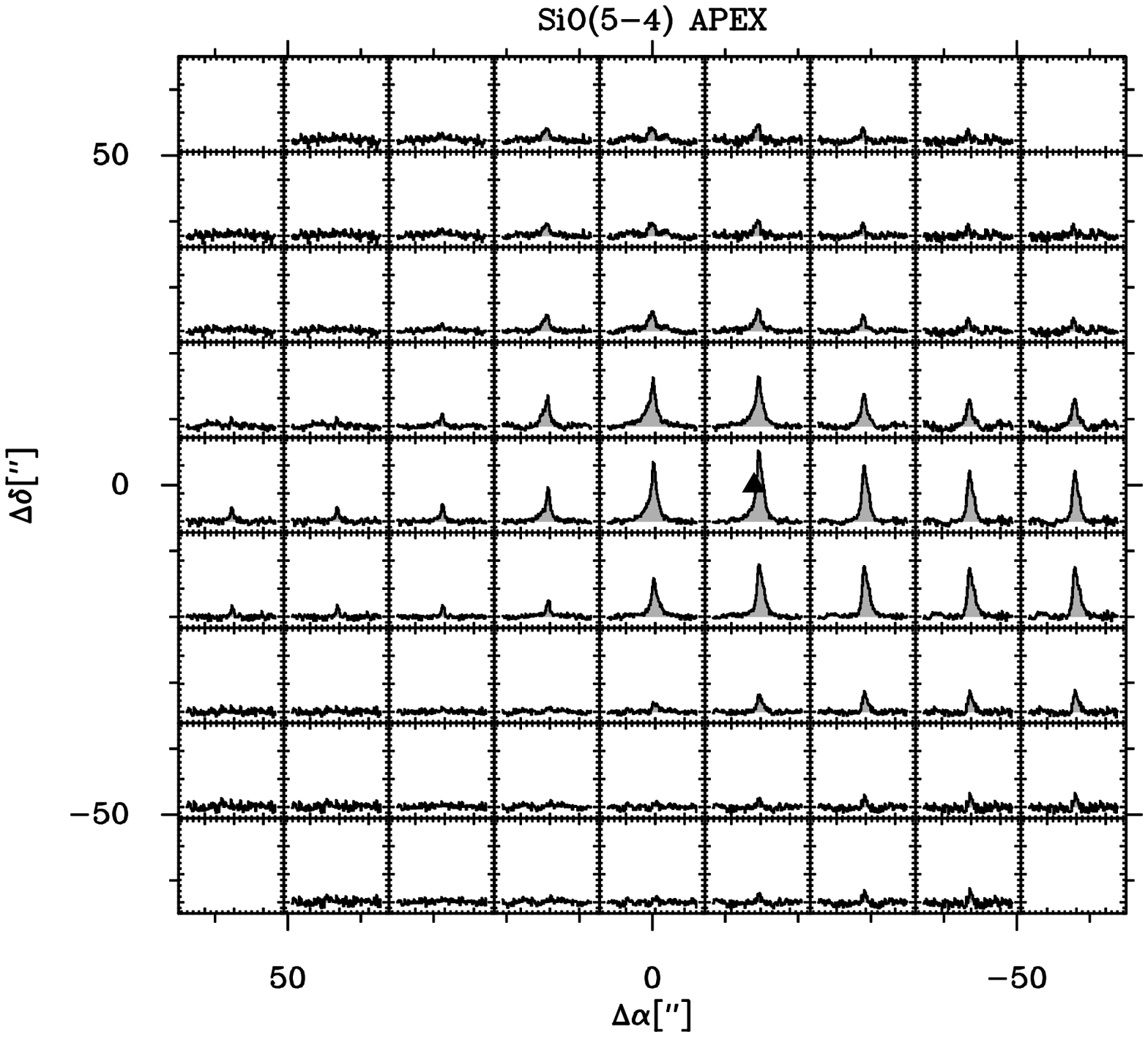}}
\subfigure{\includegraphics[angle=-90,width=8cm]{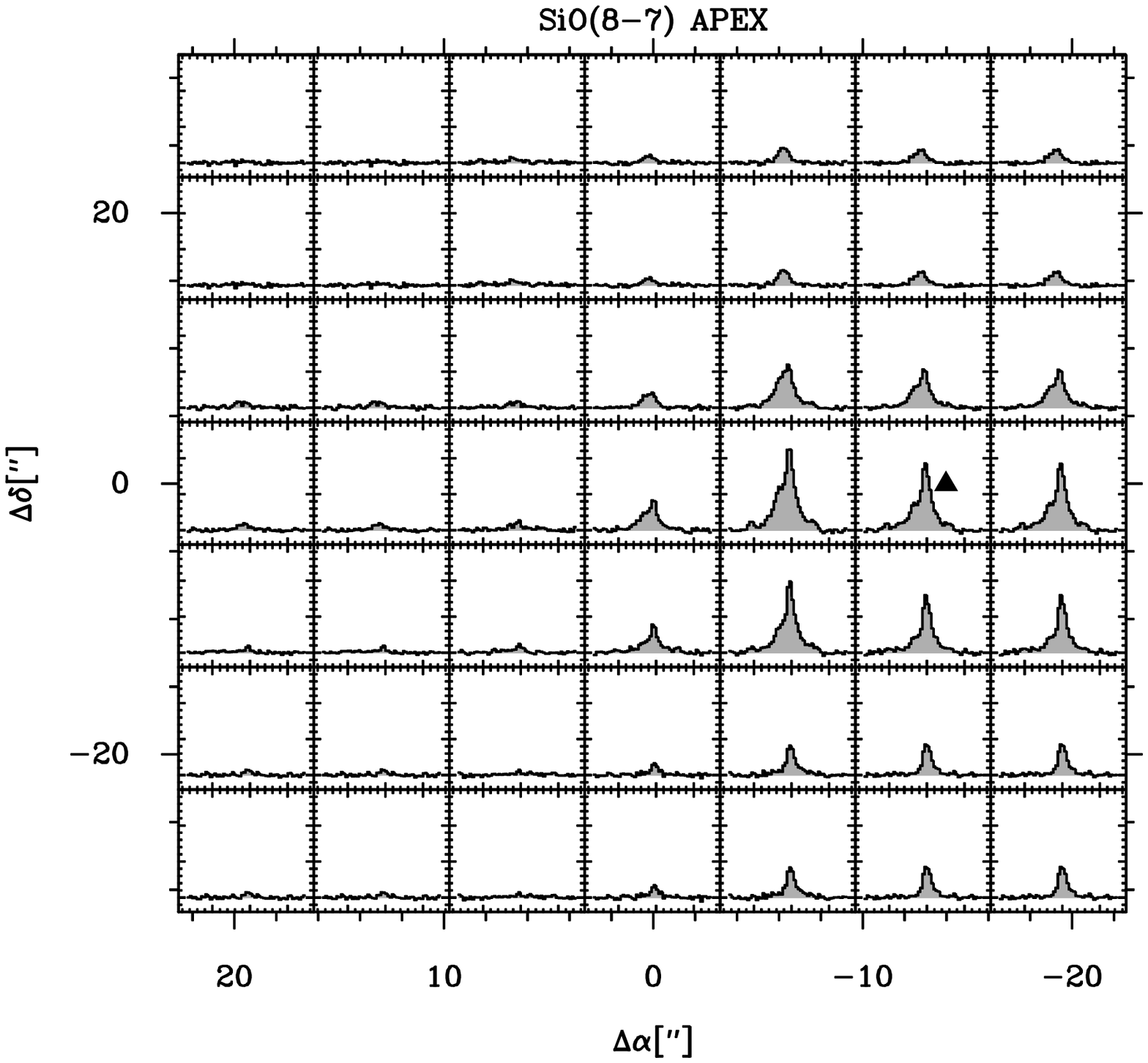}}
\caption{SiO(5--4) (top panel) and (8--7) (bottom panel) maps observed with the APEX telescope. 
The velocity axis extendes from -95 to 85~km~s$^{-1}$ for both data sets, while the temperature scale is covers -0.4 to 3~K. 
The data are shown with the original angular resolution of 30\farcs5 (SiO(5--4)) and 19\farcs1 (SiO(8--7)). 
The centre of the maps is given in Table~\ref{table_obs}. The black triangle marks the peak position (SiO-APEX, see Table~\ref{pos})
of the integrated intensity SiO(5--4) and (8--7) maps (in the velocity range $-40<\varv<40$~km~s$^{-1}$) 
smoothed to the angular resolution of the SiO(5--4) data, 30\farcs5.}\label{apexdata}
\end{figure}

\begin{figure}
\centering
 \includegraphics[angle=-90,width=9cm]{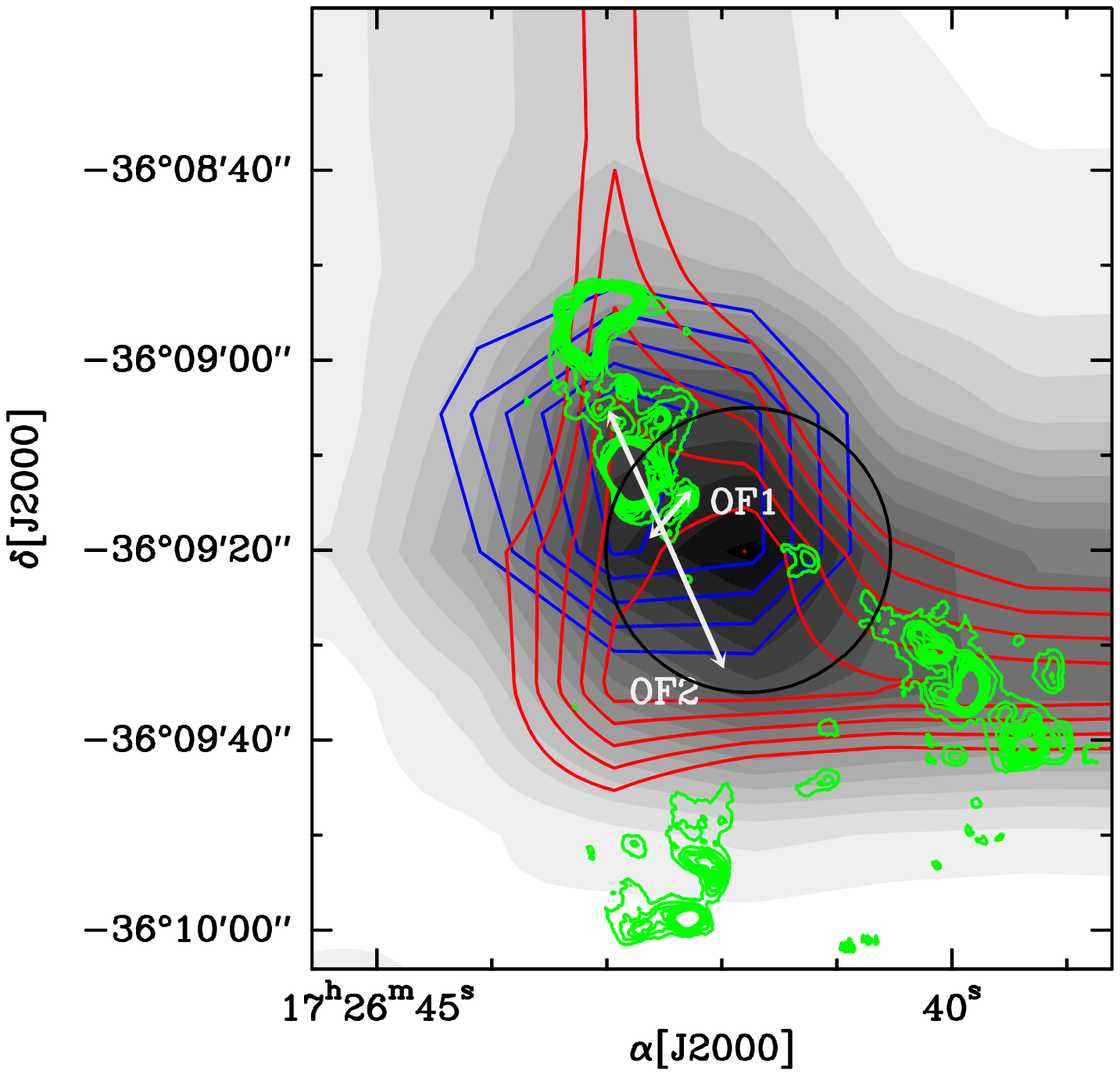}
 \caption{Total integrated intensity ($-40<\varv<40$~km~s$^{-1}$; grey shades are from 10\% of the peak emission, 
43~K~k~ms$^{-1}$, in steps of 5\%) of the SiO(5--4) transition observed with APEX.  The red and blue contours are the red- ($5<\varv<40$~km~s$^{-1}$) and blue-shifted ($-40<\varv<-5$~km~s$^{-1}$) integrated emission observed with APEX (contours are 50\% of the peak emission (9.4~K~k~ms$^{-1}$ for the red emission, 17.6~K~k~ms$^{-1}$ for the blue emission)  in steps of 10\%). The green contours represent the 
2.12$\mu$m H$_2$ emission. The white arrows label the OF1 and OF2 outflows. The black circle marks the APEX beam centred on the peak of the APEX SiO emission, as in Fig.~\ref{overview}}\label{extended}
\end{figure}

Figure~\ref{apexdata} shows the SiO(5--4) and (8--7) maps from APEX at their original angular resolution, 30\farcs5 
for SiO(5--4) and 19\farcs1 for SiO(8--7).
Given the low resolution of the APEX data, the emission from 
the different outflows in the region is not resolved.
When averaged to the common resolution of 30\farcs5, both SiO(8--7) and (5--4) maps show a maximum at  
$\alpha_{[J2000]}=17^{\rm h}26^{\rm m}41\fs77$, $\delta_{[J2000]}=-36\degr09\arcmin20\arcsec$, the centre
of the circle (representing the APEX beam) 
partially seen in Fig.~\ref{overview} (right panel). 
This position has no  counterpart in other tracers: 
it is situated at $\sim(-9\arcsec,-2\arcsec)$ from VLA-2a, 
the strongest cm source of the region, and 
at $\sim(-8\arcsec,-4\arcsec)$ from the centre of the LV SiO red-shifted emission in OF1. 
We believe, however, that the offset between SiO SMA and APEX data is due to the poor 
sampling of the APEX map. In Fig.~\ref{extended} we present the SiO(5--4) emission observed with APEX in three different
velocity ranges (total emission, red-shifted and blue-shifted emission). The map clearly shows that SiO is spread over
a large area and  is associated with several shocks from the IRAS\,17233--3606 cluster detected in H$_2$.

\begin{figure}
\centering
 \includegraphics[angle=-90,width=9cm]{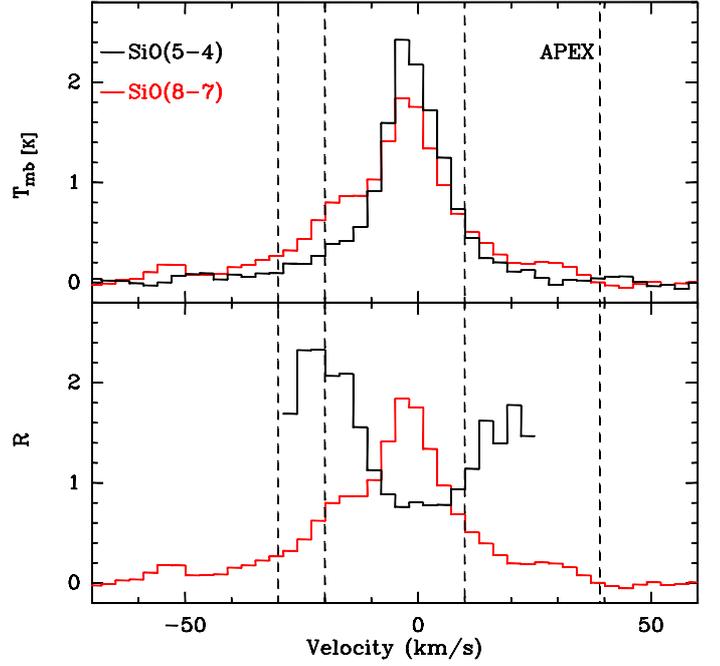}
 \caption{{\it Upper panel:} APEX spectra
of the SiO(5--4) and (8--7) lines in  the OF1 region around IRAS 17233-3606 observed at 
$\alpha_{[J2000]}=17^{\rm h}26^{\rm m}41\fs77$, $\delta_{[J2000]}=-36\degr09\arcmin20\arcsec$.
 {\it Lower panel}: 
SiO(8--7)/SiO(5--4) line ratio as function of velocity (black line). The red line is the SiO(8--7) spectrum.  The dashed lines
mark the velocity ranges used in Sect.~\ref{sio},
blue $[-30,-20]$,  and red $[+10,+39]$~km~s$^{-1}$.}\label{apex}
\end{figure}

The APEX spectra extracted at the peak of the single-dish SiO emission are shown 
in Fig.~\ref{apex}. 
Although the APEX beam at this position covers emission 
close to the ambient velocity from OF1 and OF2, the red- and the blue-shifted  
emission is dominated by OF1 both spatially and in intensity, as discussed in Sect.~\ref{zero}. 
Therefore, in the following we  assume that 
the spectra shown in Fig.~\ref{apex} are associated with OF1 only and 
represent its integrated spectra over the 30\farcs5 APEX beam. 

The SiO(8--7) line shows bright wings extending  to $\pm40$~km~s$^{-1}$, 
i.e. velocities slightly higher than those of the SiO(5--4) transition. 
Fig.~\ref{apex} also shows that the SiO(8--7)/SiO(5--4) intensity ratio 
remains almost constant ($\sim$ 0.8) at ambient velocities while it increases
 to values around 1.7 at red-shifted velocities, and 2.2 in the blue wing.  
If we assume that both SiO lines are tracing the same gas, this result 
suggests an increase of excitation conditions (discussed in Sect.~\ref{sio}) 
at high velocities, 
similar to the results obtained  towards the two low-mass Class 0 sources  L1448-mm and L1157-mm \citep{2007A&A...462..163N} through SiO observations.

\subsection{Comparison between the SMA and the APEX SiO emission}\label{zero}

We compared the SiO(5--4) APEX spectrum  at the centre of
the single-dish map with the corresponding SMA spectrum integrated over the APEX beam. The SMA total flux
corresponds to  41\% of the single-dish emission, 57\% in the
blue-wings (which dominate the spectrum at this position), 
and $\sim$ 34\% at ambient velocities. Given the broad band of the APEX and of the SMA receivers and the richness of the molecular spectrum of IRAS\,17233--3606  \citep[][ and Paper~I]{2008A&A...485..167L}, several 
spectral features  are detected simultaneously to SiO(5--4) in both datasets. For emission in the band  
from complex molecules, most likely probing the hot core, which is expected to be compact, 
fluxes between 80 and 100\% are recovered, hence confirming that the difference
between the SMA and APEX SiO spectra is due to missing flux and not to calibration issues. 

\begin{figure}
\centering
\includegraphics[bb = 196 16 551 687,clip,width=4.5cm,angle=-90]{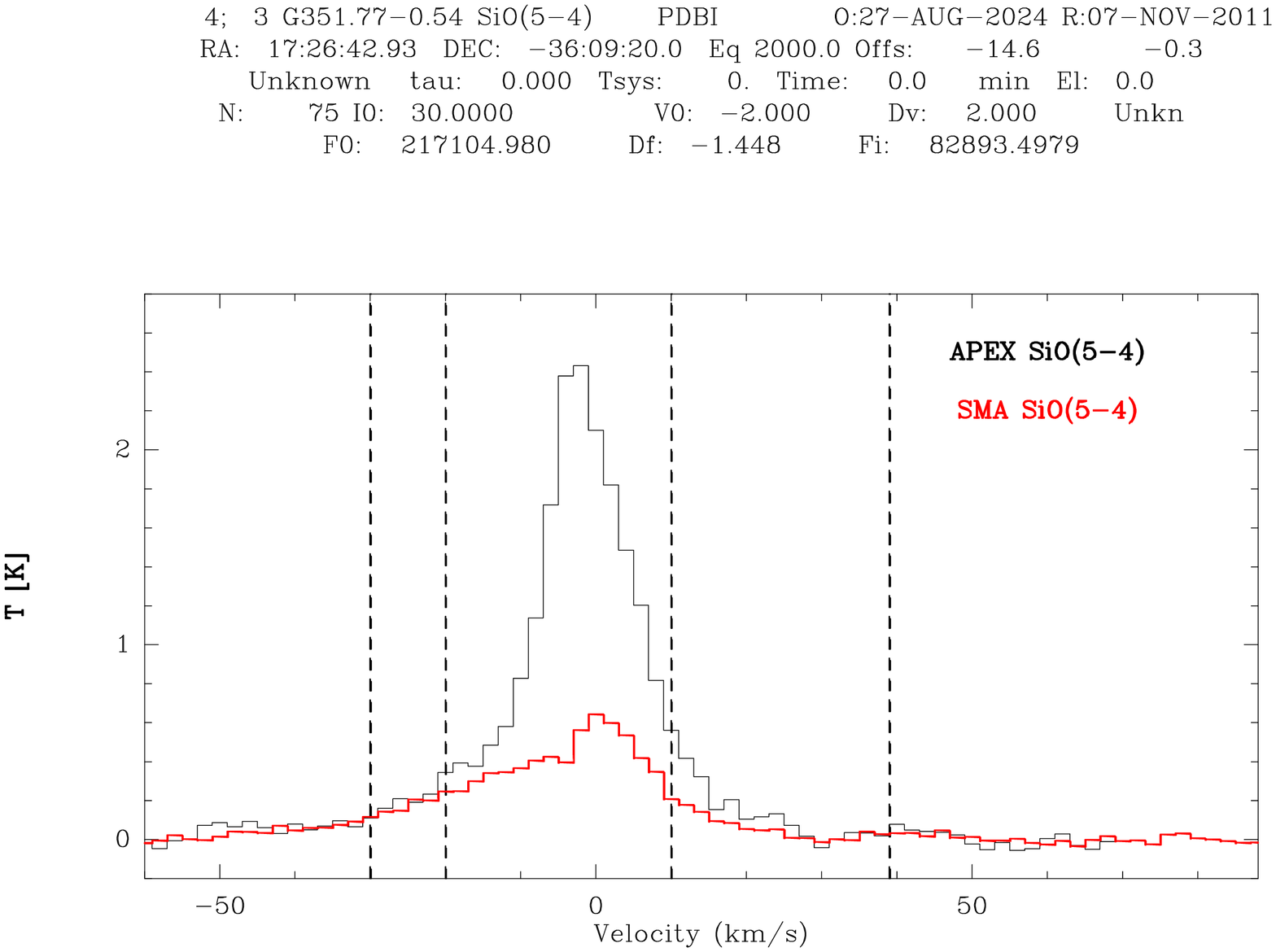}
\caption{SiO(5--4) spectra from APEX (black line) and the SMA (red line) observed towards OF1 at 
$\alpha_{[J2000]}=17^{\rm h}26^{\rm m}41\fs77$, $\delta_{[J2000]}=-36\degr09\arcmin20\arcsec$. The dashed lines
mark the velocity ranges used in Sect.~\ref{sio} for  modelling  the SiO 
emission,
blue $[-30,-20]$,  and red $[+10,+39]$~km~s$^{-1}$.}
\label{of1-apex-sma}
\end{figure}

We also compared the SMA spectrum integrated  over the APEX beam and centred on the peak of the APEX SiO emission with the corresponding APEX spectrum. The overlay of the two spectra is shown in Fig.~\ref{of1-apex-sma}.
 The SMA recovers 98\%  and  
68\% of the APEX flux in the red ([+10,+39]~km~s$^{-1}$) and blue low-velocity 
([-20,-10]~km~s$^{-1}$) ranges plotted in Fig.~\ref{overview}b. In the APEX spectrum, emission at high-velocity range ($\varv_{\rm{rd}}=[+50,+70]$~km~s$^{-1}$ and 
$\varv_{\rm{bl}}< -30$~km~s$^{-1}$) 
is not detected. However,  emission is detected at high velocities in the [-30,-20]~km~s$^{-1}$ range. In this range of velocities, the SMA recovers the whole APEX flux (Fig.~\ref{of1-apex-sma}). 
Therefore, in the range of velocities [+10,+39]~km~s$^{-1}$ and [-30,-20]~km~s$^{-1}$, the SMA data are not affected by filtering of
large structures and the morphology of Fig.~\ref{overview}c (which shows the SMA SiO integrated emission in the ranges 
[+10,+39]~km~s$^{-1}$ and [-30,-20]~km~s$^{-1}$) is reliable. 
 We show in Appendix~\ref{joint}
a joint deconvolution of the inteferometer and single-dish data performed on integrated intensity maps.
However, combining  the inteferometer data with the single-dish map is not needed for our purposes because 
the modelling of the SiO emission (see Sect.~\ref{sio}) is limited to the velocity ranges [+10,+39]~km~s$^{-1}$ and [-30,-20]~km~s$^{-1}$ where the SMA recovers the whole flux observed with APEX.

To estimate the contribution of OF1 to the APEX spectrum at the peak of the SiO emission, we computed the SMA fluxes integrated over the APEX beam centred on $\alpha_{[J2000]}=17^{\rm h}26^{\rm m}41\fs77$, $\delta_{[J2000]}=-36\degr09\arcmin20\arcsec$ and over a region enclosing OF1. The blue-shifted 
($[-30,-20]$~km~s$^{-1}$) SMA flux associated with OF1 is 66~Jy, corresponding to 66\% of the total blue-shifted flux over the APEX beam. Similarly, the red-shifted ($[+10,+39]$~km~s$^{-1}$) SMA flux in OF1 is 71~Jy (53\% of the total flux over the APEX beam). Therefore, 
OF1 contributes up to $\sim60\%$ to the integrated SiO emission observed with APEX in the velocity ranges used in Sect.~\ref{sio} for the modelling of the SiO spectra.

\section{SiO modelling}\label{sio}
In principle, specific modelling complications arise in the context of high-mass star 
formation, where the strong radiation field of the central object may affect both the physical 
and chemical structure of the shock and the excitation of molecules in the outflow region (such 
effects on the shock models are currently being studied, but their influence has not been 
properly assessed yet). The potential multiplicity of the sources or their outflows is also likely to 
generate modelling difficulties. Owing to the lack of information on the region, and as a first step towards a more complete 
description, we decided to model our SiO observations in two ways that are 
classically used to describe the SiO emission arising from shock regions generated
in  outflows of low-mass forming stars \citep[e.g., ][]{2007A&A...462..163N,Gusdorf11}. 
The first one, hereafter referred to as \lq shock-LVG', is described in Sect.~\ref{sio1} and 
is based on the combination of a shock code and a radiative transfer code based on
the large-velocity-gradient (LVG) approximation. The second one (Sect.~\ref{sio2}),
hereafter referred to as \lq slab-LVG', consists of a simpler LVG description. In
the first one, the shock is approximated by a layer of points with
varying physical and chemical conditions from one point to another. In
the second, the usual LVG physical parameters are given for one point
that epitomises the whole layer of the shock-LVG method. Although still 
distant from the probably complex \lq real' configuration of multiple shocks at work 
within the APEX beam, we believe that this two-sided approach, especially 
the shock-modelling one, is the most rigorous so far available to constrain the 
abundance of SiO and link it to the mass of the associated outflow. Indeed, 
from a computational point of view, it is still impossible to design a model 
that would account for the precise geometrical, dynamical, physical, and chemical 
structure of the region. Moreover, no observational facility can 
currently provide the very high-angular resolution necessary to separate, for instance, 
a \lq wind shock' from an \lq ambient shock' component identified in the idealistic models 
presented, e.g., by \citet{1997IAUS..182..181H}.

\subsection{Shock-LVG analysis}\label{sio1}
We first base our modelling of the SiO emission on the C-type shocks interpretation presented in details in
 \citet{Gusdorf081}. The SiO emission is accounted for
  by the combination of two codes: a state-of-the-art shock model and a radiative-transfer code based on the LVG approximation. The two codes codes are presented in detail in \citet{Gusdorf081,Gusdorf082,Gusdorf11}. We nevertheless describe the main elements relevant to our study in the following paragraph. 

The shock code calculates the dynamical, physical, and chemical structure of one-dimensional, stationary
  C-type shock models. The fractional abundances of over 125 species,
  linked by over 1000 reactions, are computed at each point of the
  shock. The gas-phase abundance of H$_2$ is calculated, based mostly
  on the treatment described in \citet{Lebourlot02}. The local
  abundance of H$_2$ is of particular importance in the context of SiO
  modelling, as it is the most abundant species to collisionally
  excite SiO. The excitation of its own 150 first levels (the most
  likely to be populated in the considered shocks) is also provided,
  making use of the H--H$_2$ collisional rate coefficients from
  \citet{Wrathmall07}. Specific grain processes, listed in
  \citet{Flower031}, are taken into account. In particular, in C-shock
  models, the  grains undergo the sputtering
  impact of neutral particles at the speed of the ambipolar
  diffusion. Because the grain cores are partly made from silicon-bearing
  material, this sputtering leads to the release of elemental Si in
  the gas phase
  (\citealt{1997A&A...322..296C,1997A&A...321..293S,May00}), where it
  reacts with O$_2$ and OH to form SiO. In the shock code, the rates
  of these two reactions are identical and correspond to the
  low-temperature measurements of the Si~+~O$_2
  \longrightarrow$~SiO~+~O reaction by
  \citet{Lepicard01}. 
 Less prominent routes to the formation of
  SiO, such as Si$^+$(OH,H)SiO$^+$(H$_2$,H)HSiO$^+$(e,H)SiO, are also
  included in our shock code.

The magnetic field parameter, $b$, defined by
  B($\mu$G)~=~$b~\times~\sqrt{n_{\rm H}~({\rm cm}^{-3})}$ where $n_{\rm H}$ is the pre-shock density, is set to 1.  
The shock velocity ranges from the minimum value likely
  to generate sputtering of the grain cores (over 25~km~s$^{-1}$), to
  the maximum value permitted by the peculiar physics of those shocks
  and by the choice of such parameters (\citealt{Flower032}).  
The age
  of the shock region can also be varied in our models, by
  varying their final flow time, that is the necessary time for a 
particle to flow from the pre-shock to the post-shock region. In practice, the final flow time is the time at which we stop the calculations.

The necessary outputs of the shock model, such as the
  temperature profile and fractional abundances of relevant species
  (SiO and colliding species) are then used as inputs of the radiative-transfer code, which also makes use of SiO spectroscopic parameters
  (Einstein coefficients, collisional rate coefficients) from \citet{2006A&A...459..297D} and \citet{2005A&A...432..369S}, and provided
  by the LAMDA database\footnote{
  http://www.strw.leidenuniv.nl/$\sim$moldata/}, to solve the
  statistical equilibrium equations and calculate the level
  populations at each point of the shock. The set of equations is
  solved thanks to the escape probability formalism. 
 The line
  emissivities and optical depth are also computed at each point of
  the shock layer for all considered transitions. Finally, we can
  compare the integrated intensity over the whole shock region, as
  resulting from the models, with the observations.

The 30\farcs5 beam of the APEX observations
    encompasses the blue- and red-lobe of OF1 (see
    Figs.~\ref{overview} and \ref{extended}, and also Fig.~\ref{apex} for our definition of
    the velocity intervals associated with these 
    lobes), which correspond to different shock regions detected at high velocities. It also
    comprises an \lq ambient velocity component' that we interpret as a consequence of the propagation of the shock wave.
    Indeed, this narrow, ambient-velocity component
    has already been seen in  bipolar
    outflows associated to low-mass forming stars, and it could
    arise either from decelerating Si-enriched post-shock gas, mixed
    with the ambient gas \citep[e.g., ][]{1998ApJ...504L.109L,1999A&A...343..585C}, or from a slow reverse C-type
    shock brought to rest by the much denser ambient medium \citep{Gusdorf081}. 
In any case, this part of the SiO emission is
    not covered by our shock modelling, which solely 
describes the more dynamical SiO emission detected in the blue- and red-wings of our observations. 
Owing to its associated velocity, this SiO emission is interpreted as arising from a shock of outflow- or jet-like origin, 
a view that is also supported by the maps shown in Figs.~\ref{zoom}, \ref{overview} and \ref{extended}.

We hence compare the red-lobe and blue-lobe integrated
  intensities of the SiO(5--4) and (8--7) lines with the results
  generated by the grid of C-shock models previously introduced in
  \citet{Gusdorf081}, for pre-shock density values of 10$^4$, 10$^5$,
  and 10$^6$~cm$^{-3}$. The observed integrated intensities are 
  $2.1\pm0.4$ and $4.4\pm0.9$~K~km~s$^{-1}$ in the blue wings and
  $3.8\pm0.8$ and $6.7\pm1.3$~K~km~s$^{-1}$ in the red wings, for
  SiO(5--4) and (8--7) respectively. The velocity ranges used in the analysis are
 $[-30,-20]$~km~s$^{-1}$  for the blue-shifted emission,  and $[+10,+39]$~km~s$^{-1}$ for the red-shifted emission as defined in Sect.~\ref{zero} and labelled in Fig.~\ref{apex}.
The error bars associated to these values 
  correspond to the 20\% calibration uncertainty intrinsic to our measurements.
  
  From our SMA maps, we derived an observed area of $\sim30$~arcsec$^2$ 
  for the SiO(5--4) emission in OF1, which corresponds to a deconvolved area of
  $\sim22$~arcsec$^2$, although the emission is unresolved along the minor axis of the emission. Therefore, we adopt 
  this value as an upper limit to the  size of the SiO emission associated with OF1. 
 A second estimate of the size of OF1 comes from the H$_2$ emission. The assumption that
SiO and H$_2$ have similar sizes in protostellar outflows is justified by high-angular resolution observations of low-mass systems \citep[e.g.,][]{2006ApJ...636L.141H,2007A&A...462L..53C}, which clearly show that SiO knots have near-infrared H$_2$ counterparts except  at high extinctions associated with dense protostellar envelope where H$_2$ is not detected. 
 For the H$_2$ counterpart of the 
  red lobe of the SiO(5--4) emission shown in Fig.~\ref{zoom}, we measure an emitting size 
  of $\sim6$~arcsec$^2$. For the blue lobe, no H$_2$ emission is detected, possibly 
  owing to extinction. However, because OF1 is seen symmetric in both SiO and SO (Fig.~\ref{zoom}), we 
  assume an emitting size of 
  $\sim6$~arcsec$^2$ for the blue-lobe, resulting in a total area $\sim12$~arcsec$^2$. Finally, we 
  assume that the size of the emitting region is the same for both SiO lines; hence, the beam-filling 
  factor is the same for both transitions, since the SiO(8--7) data were smoothed to the resolution 
  of the SiO(5--4) observations (Sect.~\ref{obs_a}). This minimises possible biases due to beam 
  dilution.
 
\begin{figure}
\centering
\includegraphics[width=7cm]{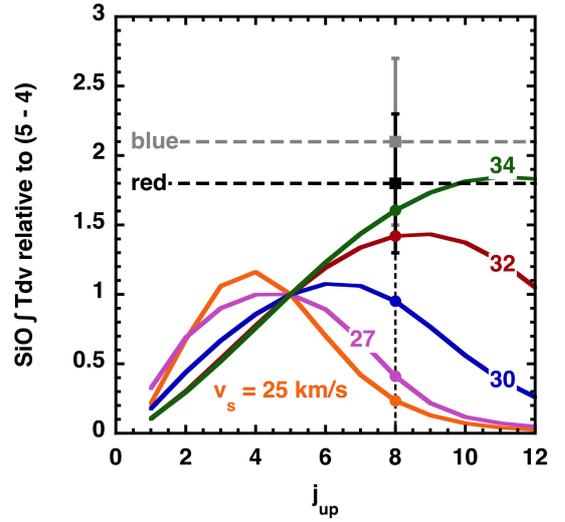}
\caption{Integrated intensities of the SiO lines relative to
      SiO(5--4). The horizontal lines show the values of the SiO(8--7)
      to (5--4) ratio, whether it is the one in the blue or red lobe
      of the outflow. The vertical line marks the position of the
      (8--7) transition on the X-axis. The colour curves provide the
      results for the fraction of our C-type shock models grid
      relevant to our study. For all these models, the pre-shock
      density is $n_{\rm H} = 10^6$~cm$^{-3}$, and the magnetic field
      parameter is $b = 1$. Only the shock velocity $\varv_{\rm s}$
      varies, and is indicated for each model on the corresponding
      line. The blue- and red-velocity range are defined in
      Fig.~\ref{apex}.}
\label{figureag}
\end{figure}

 We base our selection of a \lq satisfyingly fitting model' on two criteria. First, we use 
 the integrated intensity ratio of the SiO(8--7) to the SiO (5--4) lines. According to Fig.~\ref{figureag} 
 (adapted from Fig.~9 of \citealt{Gusdorf081}), the only model reproducing the observations is that 
 with $n_{\rm H} = 10^6$~cm$^{-3}$, $b = 1$, and $\varv_{\rm s} = 32$~km~s$^{-1}$. Below this 
 velocity, the sputtering and excitation conditions are not efficient enough, whereas above it, too 
 much SiO(8--7) emission is generated to match the observations. Then, we check that the maximum 
 line temperatures predicted by the model agree with the observed ones, which can be read 
 in main beam units in Fig.~\ref{apex}: around 0.6~K in the red and blue wings, with a maximum 
 slightly higher for SiO (8--7) than for SiO (5--4). Assuming that $\sim$60\% of the APEX SiO emission 
 originates in OF1 (Sect.~\ref{zero}), one must then account for a 
 main beam maximum temperature of $\sim$0.36~K. To convert this into brightness temperatures unit, we 
 use both the upper limit of 22~arcsec$^2$ and the H$_2$ emission size of 6~arcsec$^2$ for each 
 lobe of SiO emission. The beam size 
 of the APEX spectra being $30\farcs5$ for both transitions, the 0.36~K main beam 
 value  amounts to $\sim$12 and $\sim$44~K after correcting for dilution. This latter value agrees very decently 
 with our models predictions (Fig.~\ref{figureag}), where the line
 temperature profiles are displayed over the whole layer of the shock for the (5--4) and (8--7) transitions, 
 showing maximum temperatures of respectively 49~K and 51~K. We note that such a 
 satisfactory agreement cannot be found for other models of our grid. This  seems also to 
 suggest an emitting size for the SiO(5--4) comparable to that of H$_2$, $\sim6$~arcsec$^2$ per lobe.
 
 We can then extract the surface density corresponding to this best model, and derive two values 
 corresponding to 500 and 1000~years, which are the typical upper limits for the age of our shocks given in 
 Paper~I. We obtain  0.12 and 0.25~g cm$^{-2}$. The second value is higher than the first one because
 an older shock age implies a more-extended post-shock region, where the gas is denser. We first derive the total 
 outflow-mass comprised in the upper limit of our SiO emitting region size, $\sim22$~arcsec$^2$, adopting 
 a distance of 1~kpc. We obtain a total mass of 0.3~M$_\odot$ with the 500~years limit, and 0.6~M$_\odot$
 with the 1000~years limit. If we alternatively use the total emission region size of H$_2$, namely 
 $\sim12$~arcsec$^2$, we obtain a corresponding total outflow-mass of 0.2~M$_\odot$ for 500~yr, 
 0.3~M$_\odot$ for 1000~yr approximately.  The calculated masses 
 agree with those inferred in Paper~I (0.6--0.9~M$_\odot$) and imply a luminosity of the 
 driving source of OF1 of $\sim 10^3~L_\odot$ based on the empirical outflow mass--luminosity relation of
  \citet{2009A&A...499..811L}. This luminosity corresponds to that of a B3-type ZAMS star 
  \citep[e.g., ][]{1973AJ.....78..929P}. 
However, our estimate of the total mass of the outflow 
has to be considered as a lower limit since only a limited range of velocities was used in the analysis. 
The inferred luminosity of the powering source of OF1 is also probably  a lower limit because the  
outflow mass--luminosity relation of \citet{2009A&A...499..811L} is based on CO measurements.

\begin{figure}
\centering \includegraphics[width=9cm]{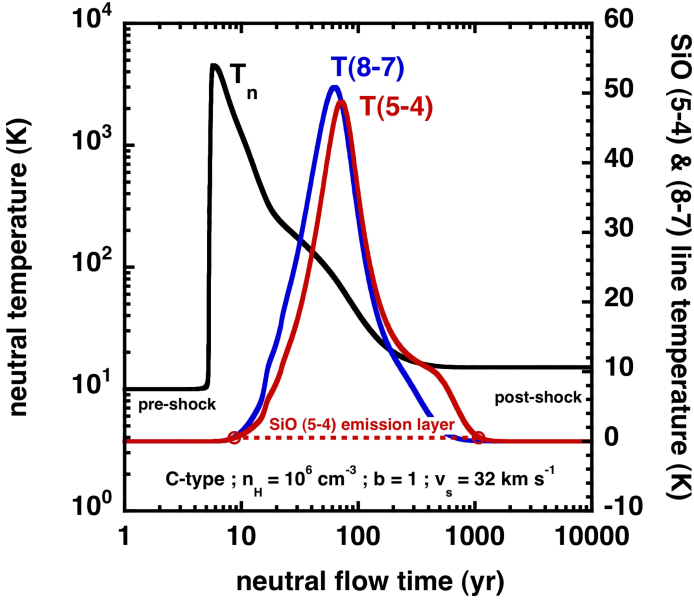}
\caption{Neutral temperature (black line) and the SiO (5--4) and
  (8--7) line temperature (red and blue lines) profiles
  over the whole shock layer, seen face-on, versus the neutral flow
  time for our best-fitting model. The characteristics of this model
  are indicated in the figure, as well as the location of pre- and
  post-shock regions. Our definition of the SiO (5--4) emission
  region (Sect.~\ref{sio2}) is also delimited by the two
  red open circles and is represented by the horizontal dashed red
  line. In this diagram, the line of sight lies on the plane of the
  figure, in the direction that goes from the pre- to the post-shock
  region.}
\label{figonline}
\end{figure}
 
\subsection{Slab-LVG analysis}\label{sio2}

\begin{figure}
\centering
\includegraphics[angle=-90,width=9cm]{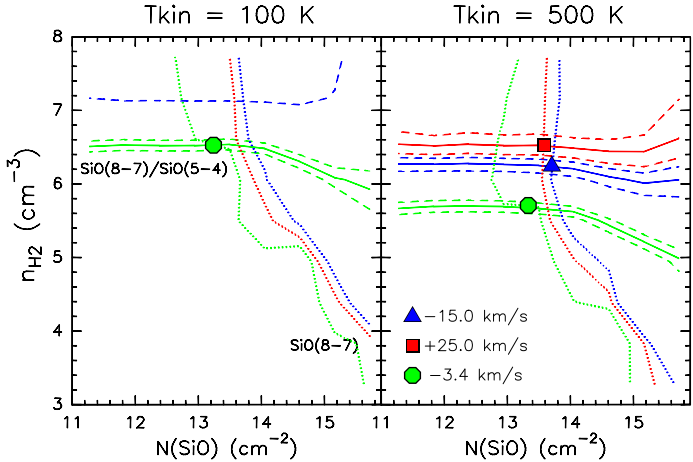}
\caption{Volume density ${\rm n_{H_2}}$ versus SiO column density,
$N$(SiO) for slab-LVG models for  $T_{\rm kin}=100$ K (left panel)
and 500 K (right). Each continuous curve
corresponds to the observed SiO(8--7)/SiO(5--4) intensity ratio
(after convolution to the same resolution, see Sect.~\ref{obs_a}),
as measured at ambient
velocity (-3.4 km s$^{-1}$, green curve) and at typical
red- (+25.0 km s$^{-1}$, red curve) and blue-shifted (--15.0 km s$^{-1}$, blue curve) velocities.
Dashed lines are for the ratio uncertainties.
The dotted curves correspond to the APEX SiO(8--7) values at the same
velocities, after correction for beam dilution using the OF1 size as derived from the SMA data.
The corresponding uncertainties are negligible
and are not shown for the sake of clarity. The symbols (circle: ambient;
square: red; triangles: blue) indicate the allowed solutions.}
\label{lvg}
\end{figure}

To verify these results and to constrain the excitation conditions of the OF1
SiO
outflow, we also performed a \lq slab' analysis of the OF1 region
using only an LVG code. We investigated the column density of SiO ($N_{\rm SiO}$) in the
range $10^{11}-10^{12}$~cm$^{-2}$,
H$_2$ densities ($n_{\rm H_2}$) of $10^3-10^8$~cm$^{-3}$ and kinetic temperatures ($T_{\rm kin}$)
of 10--600~K.
Figure
\ref{lvg} shows our typical model results for a kinetic temperature
of 100 and 500 K.
We analysed the line intensities at LSR velocity as
well as at the average velocities of the blue- (--25 km s$^{-1}$) and
red-shifted (+15 km s$^{-1}$) lobes. Figure \ref{lvg} indicates temperatures higher than 100 K
for
the red and blue lobes: lower
temperatures would imply unreasonable densities higher than 10$^{8}$
cm$^{-3}$. The $n_{\rm H_2}$ values for $T_{\rm kin}$ = 500 K are $\simeq$
5 10$^{6}$ cm$^{-3}$.
Since this is very likely an upper limit to the kinetic temperature of the gas, these densities have
to
be considered as lower limits. The inferred SiO column densities are $\sim 0.5-
1~10^{14}$ cm$^{-2}$ 
for both lobes. On the other hand, at ambient velocity we can
infer $N_{\rm SiO}$ of $2-5~10^{13}$ cm$^{-2}$ and densities higher
than 10$^{5}$ cm$^{-3}$ without any clear constraint on kinetic
temperature.

For consistency purposes, we cross-checked the results yielded by our
two modelling methods, shock-LVG vs. slab-LVG. To do so, one must return to the shock models and extract the ranges of variation over the
shock layer of the parameters, which can only then be compared with the
single-point values of the same parameters, which prevail in the slab-LVG analysis.
The shock code outputs consist of a series of profiles
    of physical and chemical quantities (e.g., temperature, abundance)
    calculated over the shock layer. Our LVG code, when used in
    combination with the shock code, generates a series of outputs in
    the form of similar profiles of emissivities quantities (e.g., level
    populations, line emissivities). For the model that was chosen
    to describe the OF1 observations, Fig.~\ref{figonline} shows the
    neutral temperature and  the local emissivity of the SiO
    (5--4) line over the whole shock layer, which is represented on the
    x-axis by means of the neutral flow time of the shock, i.e. the
    necessary time for a neutral particle to fly through the shock
    region. This representation allows us to empirically define the
    SiO (5--4) emission region as the region where the
    emissivity of the line is significant. By significant we  mean
    that the emission region consists of all the points where the
    emissivity of the line exceeds 1\% of the maximum line
    temperature, that is about 0.5 K for the considered model. We
    then compare the range of variations over this emission region, of
    the quantities that are also used as input parameters in the
    slab-LVG method. This way, we performed a proper comparison between the range-values inferred from the shock-LVG analysis and the single-values yielded by the slab-LVG method.
    These are reported in 
    Table~\ref{table1}. In our shock-LVG models, only one parameter is
    not expressed in the form of a range of values: the SiO column
    density because it corresponds to the SiO density integrated over the whole shock layer.
 Note that the definition of the
    LVG parameter is the same in both the slab-LVG and shock-LVG
    cases, but that its expression differs. For the slab-LVG modelling, it is simply equal to $N_{\rm SiO}$ (cm$^{-2}$)~/~$\Delta \varv$
    (km s$^{-1}$), which is equivalent to the shock context
    formulation $10^5~\times~n$(SiO)(cm$^{-3}$)~/~grad\_$\varv$ (s$^{-1}$),
    where grad\_$\varv$ is the velocity gradient, and where the $10^5$
    factor only arises because of the conversion from km to cm.

\begin{table*}
\caption{Properties of SiO-emission regions inferred from face-on
  C-type shock models used in combination with an LVG code, and from
  homogeneous slab-LVG models.}
\label{table1}      
\centering
\begin{tabular}{c c c c c}        
\hline medium \hspace{1.2cm} & \multicolumn{3}{c}{\textbf{slab-LVG }}
& \hspace{.7cm} \textbf{shock-LVG} \\ properties \hspace{1.2cm} & red
lobe & blue lobe & ambient & \hspace{.7cm} best model \\ \hline \hline
$T_{\rm kin}$ (K) \hspace{1.2cm} & 100--500 & 100--500 & n.c. & \hspace{1cm} 15--1745
\\ $n{\rm_{H_2}}$ (cm$^{-3}$) \hspace{1.2cm} & $\ge 5\times 10^6$ & $\ge 5\times
10^6$ & $\ge 10^5$ & \hspace{.7cm} 3.3$\times 10^6$--2.4$\times 10^7$
\\ $N_{\rm SiO}$ (cm$^{-2}$) \hspace{1.2cm} & 5$\times 10^{13}$ &
$10^{14}$ & 5$\times 10^{13}$ & \hspace{.7cm} 2.4$\times 10^{14}$
\\ LVG parameter \hspace{1.2cm} & 2.5$\times 10^{12}$ & 5$\times
10^{12}$ & 2.5$\times 10^{12}$ & \hspace{.7cm} 6.4$\times
10^{11}$--5.0$\times 10^{14}$ \\ \hline
\end{tabular}
\end{table*}

The results  shown in Table~\ref{table1} are
    remarkably consistent and seem to support the view that tools that have 
    been proved efficient in the context of low-mass star formation can be used 
    in higher mass cases, with slightly increased densities. However, our best-fitting
    shock model was only approaching the observed SiO line ratios, and despite 
    additional investigations in the parameter space, we were unable find a model likely to 
    match (8--7) / (5--4) ratio values as high as 2, for instance. If confirmed as 
    characteristic of massive star formation processes, such a high ratio could be 
    accounted for by considering grain-grain interactions within the shock layer. Such 
    interactions have been described as important at high pre-shock densities, and 
    their feedback effects on both the chemical and dynamical structure of shocks have
    been studied by \citet{Guillet07,Guillet09,Guillet11}. Unfortunately, those authors 
    could not generate any self-consistent solution for the pre-shock densities at work 
    here, namely for $n_{\rm H}$ = 10$^{6}$ cm$^{-3}$. Alternatively, the influence of a 
    strong radiation field generated by the driving source of the outflow both on the 
    shock structure itself and on the corresponding molecular excitation might
    yield high SiO line ratios. The rather complex implementation of such processes 
    in shock models is work in progress.
    The present analysis indicates volume densities
    higher than 10$^5$ cm$^{-3}$ (see Table 3).
    \citet{2007A&A...462..163N} performed an LVG analysis of SiO emission towards
    the nearby prototypical Class 0 low-mass objects L1148 and L1157,
    using single-dish  data on angular scales of 10$\arcsec$.
    No tight constraints have been obtained for kinetic temperature,
    whereas the volume densities were found to be between 10$^5$ and 10$^6$ cm$^{-3}$.
    In addition, \citet{Gusdorf081} modelled the SiO emission from L1157
    using a C-shock code leading to pre-shocked densities of
    10$^4$--10$^5$ cm$^{-3}$. Thus, although completely different spatial scales
    are involved, the volume densities here inferred for IRAS\,17233--3606 
    appear to be consistent with those derived for L1448 and L1157.

Although the results from the slab LVG approach are consistent with
those inferred through the more rigorous shock-LVG analysis, we stress here the importance of shock models for deriving physical
parameters such as the SiO abundance and therefore the mass of the
outflow: while in the slab-LVG method one has to rely on assumed
values of $X_{\rm{SiO}}$, $X_{\rm{SiO}}$ is consistently computed in
the shock model, thus decreasing the uncertainties in the estimate of
the outflow mass.

\section{The nature of OF1}

 The present data sets  trace the outflow activity driven by the YSOs in the cluster, and contribute to 
draw a clearer picture of the region than that obtained by our previous CO data (Paper~I).
 Our previous observations of OF1 were limited to EHV red-shifted CO(2--1) emission and to the detection of H$_2$ emission also constrained to only the red lobe of OF1.  The bipolar nature of OF1 was suggested only by the SO data and by water maser spots. Our present SMA data confirm 
that the emission associated with OF1 is due to a bipolar collimated outflow  well traced by SiO in both red- and blue-shifted emission. 
This is consistent with findings for  jets driven by Sun-like protostars, where it is possible to see
that CO traces not only the jet but
also the walls of
the cavity opened by the jet itself (e.g., \citealt[][ their Fig.~5]{2007ApJ...670.1188L}, and \citealt[][ their Fig. 1]{2012A&A...548L...2C}).
On the other hand, 
SiO emission is usually associated with primary jets in the case of 
low-mass YSOs \citep[e.g.,][]{2006ApJ...636L.141H,2012A&A...548L...2C} and 
generally suffers minimal contamination from infalling envelopes or swept-up 
cavities. 

Our data strongly suggest
that the emission associated with OF1 traces a bipolar jet from 
the IRAS\,17233--3606 cluster.  
This view is additionally  supported by the high velocities detected in SiO, and by the collimated morphology 
of OF1, which is $\sim 6$ according to the H$_2$ images,  
among the largest ever reported in high-mass star
formation \citep[see for example][]{2002A&A...383..892B,2004ApJ...608..330B,2004ApJ...604..258S,2011ApJ...735...64W}. This value corresponds to the ratio between 
the major semi-axis and the minor semi-axis of OF1 as derived from the H$_2$ map. Additionally, 
the high temperature ($T>100$~K) derived through the SiO analysis is  suggestive
of material very close to the primary jet.

Although the driving source of OF1 cannot be identified yet given the 
still relatively low resolution of the SMA 
data ($\sim3000$~AU), the best candidate source is 
VLA-2a, which is the strongest cm-continuum source in the region and  
 also the compact source closest to 
the H$_2$ jet-like feature. Alternatively, VLA-2b could also be a possible candidate for 
the powering source of OF1.
In principle, we 
cannot rule out the hypothesis that the driving source of the observed
SiO bipolar emission is a low-mass YSO in the cluster. However, 
there is some 
indirect evidence that OF1 is indeed driven by a massive YSO. First, we determined  the total mass of OF1  
through the shock modelling of SiO emission 
without any assumption of the SiO abundance (Sect.~\ref{sio1}) and using the area inferred from the H$_2$ emission.
From this mass, we inferred a  luminosity of  $\ge10^3$~L$_\odot$ for  the OF1 driving source, which   agrees with an earlier than B3-type star. 
The continuum flux of VLA-2a in the cm \citep{2008AJ....136.1455Z} 
is also compatible with that of an early B ZAMS
  star. The cm continuum flux of VLA-2b \citep{2008AJ....136.1455Z} is also compatible with an early B-type star, although less massive than VLA-2a (a B3 ZAMS star compared to a B1 ZAMS star).
To our knowledge, this is one of the first indirect evidences of a SiO collimated bipolar outflow from an early B type YSO, although only higher
spatial resolution maps will allow a direct proof of its existence. 
Previous studies reported red- and blue-shifted SiO emission from massive star-forming regions 
\citep{1999A&A...345..949C,1999AJ....118..477H,2007ApJ...654..361Q,2007A&A...470..269Z,2013A&A...550A..81C}. 
Among these, IRAS\,20126+4104 \citep[e.g.,][]{1999A&A...345..949C,2008A&A...485..137C} probably represents the best
example of a jet from a massive YSO, since the nature of the jet is proved by SiO, H$_2$ and [Fe{\sc~ii}] observations. 
However, the mass of the powering source is $\sim 7$~M$_\odot$ \citep{2005A&A...434.1039C}.
In one case \citep[IRAS\,18264--1152,][]{2007ApJ...654..361Q} SiO is also detected
at high velocities, as in the present case of OF1.  However, the difference between these studies and our case 
is that  several features typical of low-mass jets (collimation, emission at high and extremely-high velocity, the parameters of the gas derived by the shock model) are all found in OF1 while the other examples show only some of these properties.

\section{Conclusions}\label{con}
We reported and analysed SMA and APEX SiO observations of the IRAS\,17233--3606 star-forming region.
The SMA data allowed us to clearly resolve the OF1 outflow with a resolution of $\sim3000$~AU and identify 
its SiO emission as due to a collimated bipolar outflow associated
with EHV CO, H$_2$ and SO. 

The OF1 SiO(5--4) and (8--7) APEX spectra suggest 
an increase of the excitation conditions at high velocities and point to hot gas close to the primary jet, as found
in outflows from low-mass YSOs. 
From the shock-LVG analysis of the APEX spectra, 
we derived a mass of $>0.3~$M$_\odot$ for OF1, which  implies a 
luminosity L$\ge 10^3$~L$_\odot$ for the OF1 driving source.
Although the driving source of OF1 cannot be yet identified at the resolution of our current SMA data, 
the best candidate source,VLA-2a, has a continuum flux in the cm 
that also agrees with that of a ZAMS star with a luminosity of a few 10$^3$~L$_\odot$.

 The high angular resolution spectroscopic observations that we collected on OF1 in IRAS\,17233--3606 support a picture similar to that seen in studies of jets and outflows from low-mass YSOs.  The SiO emission at LV peaks closer to
the driving source
than the SiO-HV peak; moreover, SiO traces a region of the outflow/jet system heavily obscured
in the H$_2$ image. Finally, the SiO and SO molecules have a very similar distribution as found in collimated outflows driven by low-mass Class 0 YSOs.

In terms of shock modelling,
  we found it possible to model SiO emission in shock
  regions generated by high-mass protostars with the same tools developed  for shocks from low-mass YSOs. The shock parameters inferred by the SiO modelling are compatible with those 
found for low-mass protostars, only with higher
  pre-shock densities. The same conclusions arise from a simpler
  LVG slab analysis.

\begin{acknowledgements}
The authors would  like to thank the anonymous referee who helped to clarify the results and improve the discussion. 
A. Gusdorf acknowledges support by the grant ANR-09-BLAN-0231-01 from the French {\it Agence Nationale de la Recherche} as part of the SCHISM project. 
\end{acknowledgements}

\Online

\begin{appendix}
\section{Joint deconvolution of the SMA and APEX data}\label{joint} 
\begin{figure}
\centering
\includegraphics[width=10cm,angle=-90]{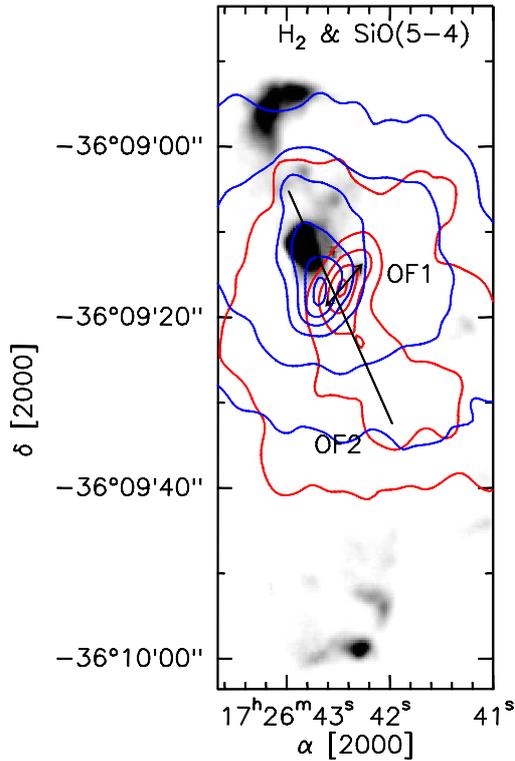}
\caption{Integrated emission
of the  SiO(5--4) line in the SMA+APEX combined data. 
The blue and red contours are the  blue- 
and red-shifted emission
 ($\varv_{\rm{bl}}=[-50,-20]$~km~s$^{-1}$, from 10$\sigma$ (1 Jy/beam) in steps of 10$\sigma$; 
$\varv_{\rm{rd}}=[+10,+39]$~km~s$^{-1}$, from 10$\sigma$ (1 Jy/beam) in steps of 10$\sigma$). The grey image represents the  H$_2$ emission at 2.12$\mu$m.}\label{combined}
\end{figure}

We performed a joint deconvolution of the SMA and APEX observations in
the image plane using the task feather in the Common Astronomy
Software Applications \citep[CASA,][]{2007ASPC..376..127M}.  A snapshot
of the results of joint deconvolution of SMA and APEX data is shown in
Fig.~\ref{combined}. While the large-scale details are filtered out by
the interferometer, the combined image restores the large-scale
information. i.e. {\it(1)} the combined image faithfully reproduces the
large-scale (extended) structure, {\it(2)} the joint deconvolution image
also conserved the total flux as observed with the single-dish. We
tested this by smoothing the combined image to the resolution of the
single dish beam. We also attempted to combine the data sets in the
UV plane. However, the flux calibration factor (ratio of single-dish
to interferometer flux in the overlapping UV space) was unexpectedly
high.  Although we could
not ascertain the reason for the large scaling factor, the smaller area mapped with the single dish 
relative to SMA area might
contribute to the failed combination.

In Sect.~\ref{zero}, we demostrated
that the SMA recovers the whole flux observed with APEX in the
velocity ranges used in the analysis in Sect.~\ref{sio} and therefore the combination of 
the two datasets was not needed  for our purposes.

\end{appendix}

\end{document}